\def\bq{\begin{equation}}
\def\eq{\end{equation}}
\def\bqa{\begin{eqnarray}}
\def\eqa{\end{eqnarray}}
\def\bqb{\begin{eqnarray*}}
\def\eqb{\end{eqnarray*}}
\documentstyle[12pt]{article}
\def\pr#1#2#3{ Phys. Rev. ${\bf{#1}}$, #2 (#3) }
\def\prl#1#2#3{ Phys. Rev. Lett. ${\bf{#1}}$, #2 (#3) }
\def\pl#1#2#3{ Phys. Lett. ${\bf{#1}}$, #2 (#3) }

\def\np#1#2#3{ Nucl. Phys. ${\bf{#1}}$, #2 (#3) }
\def\zp#1#2#3{ Z. Phys. ${\bf{#1}}$, #2 (#3) }

\def\prep#1#2#3{ Phys. Rep. ${\bf{C#1}}$, #2 (#3) }

\global\nulldelimiterspace = 0pt

\def\roughly#1{\mathrel{\raise.3ex
    \hbox{$#1$\kern-.75em\lower1ex\hbox{$\sim$}}}}

\def\bq{\begin{equation}}
\def\eq{\end{equation}}
\def\bqa{\begin{eqnarray}}
\def\eqa{\end{eqnarray}}
\def\bqb{\begin{eqnarray*}}
\def\eqb{\end{eqnarray*}}

\def\cot{\cos\theta}

\begin{document}
\pagenumbering{arabic}
\thispagestyle{empty}

\thispagestyle{empty}

\vspace*{1cm}
\hspace {-0.8cm} PM/94-04 \\
\hspace {-0.8cm} UTS-DFT-94-02\vspace {2cm}
\begin{center}
{\Large\bf Information and Discrimination  }
\vspace{0.5cm}\\
{\Large\bf from b Quark Production on Z Resonance  }
    \vspace{1cm}  \\
 {\Large D.Comelli, C. Verzegnassi
\footnote {Work partially supported by NATO.}}\\
 Dipartimento di Fisica Teorica,
 Universit\`{a} di Trieste \\
 Strada Costiera, 11 Miramare, I-34014 Trieste, \\
 and INFN, Sezione di Trieste, Italy
\vspace {1cm}\\
 {\Large F.M. Renard }
  \\
 Physique
Math\'{e}matique et Th\'{e}orique,\\
CNRS-URA 768, Universit\'{e} Montpellier II \\
 F-34095 Montpellier Cedex 5
\vspace {1.5cm}  \\
 {\bf Abstract}
\end{center}
\noindent
We introduce and define operatively  in a  model independent way
a new ``heavy"
b-vertex
 parameter, $\eta_b$, that can be derived from the measurement of
a special polarization asymmetry for production of b-quarks on Z
resonance. We show that the combination of the measurement of $\eta_b$
with that of a second and previously defined ``heavy" b-vertex
parameter $\delta_{bV}$ can discriminate a number of models of
New Physics that remain associated to different ``trajectories" in the
plane of the variations of the two parameters. This is shown in
particular for some popular SUSY and technicolor-type models. In
general, this discrimination is possible if a measurement of
\underline{both} parameters is performed.

\setcounter{page}{0}
\newpage

\baselineskip=24pt

\section{Introduction}
 In the first four years of running at LEP1, a remarkable experimental
effort has allowed to collect a number of events that begins to
approach the $10^7$ limit, that was once considered   as nothing
more than an optimistic dream.
This is the result of a number of machines's modifications or
improvements, whose main features can be found in several recent
publications or in the Proceedings of dedicated Workshops .\par
 Meanwhile, on the other side, the theoretical approach to
the interpretation of this huge amount of data has also been adapted
and improved. In fact, in very recent years it has become clear that,
to a certain extent, the comparison of the various results with the
Minimal Standard Model (MSM) predictions, and the consequent search of
possible signals of New Physics through small deviations due to
one-loop effects, can be performed in a rigorously model-independent
way. In particular, it has been stressed [1] that the
\underline{leptonic} charged processes can be ``read'' in terms of two
parameters, originally called $\epsilon_{1,3}$ in ref.[1], in a
totally unbiased way, that is for models of New Physics that are
willing, or able, to modify any of the three classes (self-energies,
vertices, boxes) of one-loop radiative effects (in practice, owing to
their intrinsic irrelevance for LEP 1 Physics
at the starting MSM level, boxes are
usually neglected for this kind of search).\par
 The generalization of the previous philosophy to \underline{hadron}
production requires some preliminary choice. In fact, the extra vertex
corrections that enter the theoretical expressions are not universal
and introduce new unwanted degrees of freedom of both  ``light" (in
practice, massless)  and  ``heavy" quark type. The latter effect
is, for the specific case of $e^+e^-$ Physics on Z resonance, entirely
due, in the MSM, to that component of the $Zb \bar b$ vertex due to
the charged would-be Goldstone exchange that behaves as $m_t^2$ for
large top masses, as it has been exhaustively shown in the literature
[2]. Since various models of New Physics generally contribute either
the light quark and lepton or the heavy quark degrees of freedom but
not both, it becomes necessary to develop an appropriate  strategy
 to perform
a satisfactory search of New Physics effects.\par
 A first possible attitude is that of only considering those models
that would \underline{not} contribute the lepton and light quark
vertices. Then, one only has to add
 to the ``canonical" quantities $\epsilon_{1,3}$  one
extra parameter . For the latter, an operational
definition should now be provided. The original proposal [3], [4], to
which we shall stick in this paper, was to define the vertex
correction $\delta_{bV}$ from the ratio of the $Zb \bar b$ and $Zs
\bar s$ partial widths i.e.
\bq  {\Gamma_b \over \Gamma_s} \equiv 1 + \delta_{bV}
\eq
where the physical b width (we follow in fact the slightly modified
version given in ref.[4]) should be taken.\par
 Once the definition eq.(1) is chosen, a systematic analysis of all
LEP 1 data that includes both leptonic and hadronic channels can be
performed in terms of three parameters e.g. $\epsilon_1$,
$\epsilon_3$, $\delta_{bV}$ or $\Delta \rho$, $\Delta_{3Q}$,
$\delta_{bV}$ in the notation of ref.[4], for the previously selected
set of models of New Physics.
This was proposed in ref.[4] and also in another series of papers [5],
where an essentially similar $Zb\bar b$ vertex parameter was introduced
(and defined $\epsilon_b$). Without entering the details of the
methods, it should be stressed that the parameter $\delta_{bV}$ as
defined in eq.(1) is operationally connected to the experimentally
measured ratio $R_b = {\Gamma_b \over \Gamma_h} $ by
 the relation (valid in the considered class of
models)
\bq  {\Gamma_b \over \Gamma_h} \equiv R_b = \frac{13}{59}(1+
\frac{46}{59}
\delta_{bV} -\frac{23}{59}(\delta_1-\delta_2) +
\frac{2}{65} \Delta \kappa '+0.1\frac{\alpha_s(M^2_Z)}{\pi}+
``negligible'')
\eq
 Here $\Delta \kappa '$ is a radiative correction entirely fixed
 by the measurements
at LEP1 (SLC) of the effective angle $s^2_{EFF}(M^2_Z)$
 (which can be identified for
practical purposes with each of the existing popular definitions [6])
\bq s^2_{EFF}(M^2_Z) = s^2(1+\Delta \kappa ') \ \,\ \ s^2 \simeq0.231
\eq
and the weight of $\alpha_s(M^2_Z)$ is practically irrelevant. The
 parameters $\delta_{1,2}$ are
certain combinations of leptonic and \underline{light} quark vertices,
whose (small ) numerical value can be exactly computed in the MSM;
their definition has been given in a previous paper [7], to whose
notations we shall stick. Thus,
if New Physics does not affect the light fermion vertices,
$R_b$ can provide the unbiased value of $\delta_{bV}$, to be compared
with the MSM prediction.\par In fact, an overall analysis of data
 is more
elaborated and includes other variables as well. The full details can
be found in refs.[4] and [5]; the point that we want to stress here is
that, after the most recent LEP1 communicated data [8], this type of
investigation leads to the conclusion that $\epsilon_1$,
$\epsilon_3$, (or $\Delta \rho$, $\Delta_{3Q}$  in the notation
of ref.[4]) are now perfectly consistent with the MSM predictions.
This means that the small discrepancy that might have been present in
the previous determinations
 of $\epsilon_3$ ($\Delta_{3Q})$ has now been (almost)
 completely washed
out. On the contrary, the possibility of a \underline{small} deviation
is still allowed in the heavy vertex parameter $\delta_{bV}$,
 since one has
now [9]:
\bq \delta_{bV} =(-12 \pm 10) 10^{-3}
\eq
 and the MSM tolerance region (corresponding to the last bound
$m_t \geq 113$ GeV [10]) is
\bq \delta^{MSM}_{bV} \leq -0.016
\eq
 One possible question that becomes relevant at this stage is
whether the assumption that  the light fermion vertices remain
unaffected has some  experimental support. To answer this question one
should identify (at least) one quantity that is only reacting to such
kind of New Physics effect. In fact, this ``light vertex indicator" has
been proposed in ref.[7] as a certain combination of hadronic and
leptonic widths and of $ s^2_{EFF}(M^2_Z)$, and defined D.
 At one loop, it is only
affected by a certain combination of light fermion vertices parameters
(different from that entering $R_b$ eq.(4)). For that combination, the
experimental data show a very good agreement with the MSM predictions,
 as
fully discussed in ref.[7].\par
 If one believes that a small discrepancy is still present in $R_b$
eq.(2),  two attitudes become possible. One is that of addressing
the full responsability to the heavy b vertex parameter $\delta_{bV}$.
 The other
one is that of thinking that an effect of the light vertex type could
modify the combination entering $R_b$ (with $\delta_{bV}$
 unaffected), but not
that contained in D. Although a priori no possibility should be
 discarded,
we feel that the second choice appears somehow unnatural. Therefore, we
shall first concentrate on the more plausible solution, in which
 New Physics
only  affects $\delta_{bV}$ as a direct consequence of the fact
 that the b quark is,
for a certain type of effects, to be considered as a member of a
``heavy" doublet.\par
In terms of shifts in the (conventionally defined) vector and axial
vector $Zb\bar b$ couplings, the effect of New Physics on
 $\delta_{bV}$ can be
parametrized as
\bq \delta^{NP}_{bV} = -\frac{4}{1+b^2}[b \; \delta g^H_{Vb} +
 \delta g^H_{Ab}]
\eq
 where
\bq  b= 1- \frac{4}{3} s^2
\eq
 and $ s^2 $ is defined by
 eq.(3).  The subscript ``H" denotes the fact that we are now
considering ``heavy" quark type of effects.\par
For the purposes of our search, it would be extremely useful to define
and to measure a certain experimental quantity where a different
combination of shifts in $g_{Vb}, g_{Ab}$ enters. In fact,
 such a quantity exists
and has been proposed a few years ago [11]. It was defined as the
``longitudinally polarized forward-backward $b \bar b$ asymmetry" and
usually called $A_b$
\bq
A_b = \frac{ \sigma (e^-_L \rightarrow b_F) -\sigma (e^-_R \rightarrow
 b_F)-
\sigma (e^-_L \rightarrow b_B)+\sigma (e^-_R \rightarrow b_B)}
{\sigma (e^-_L \rightarrow b_F)+\sigma (e^-_R \rightarrow b_F)
 +\sigma (e^-_L \rightarrow b_B) +\sigma (e^-_R \rightarrow b_B)}
\eq
 and, as one sees, it requires the availability of longitudinally
polarized electron beams. The remarkable feature of $ A_b $
is that of only depending on the couplings of Z to b,
as it was  stressed in Ref.[11] .  This explains
the great  potential interest of its measurement, that will
be performed in a very near
future at SLC if the very encouraging trend of recent progress in the
machine performance is (hopefully) going to continue [12], and might
also  be performed in a not too far future at LEP if a phase with
 polarized
beams became operative [13]. If this were the case,  an
extremely fruitful combination with the results on $R_b$ obtained by
unpolarized measurements at LEP1 would become possible, which could
allow  to draw unexpected   conclusions on this fascinating and
still existing possibility of small MSM failures.\par
This short paper is dedicated to the study and to the
exploitation of the possible theoretical consequences
of a combined determination of $ R_b $ and $ A_b $.
 In Section 2, we shall very
briefly recall the needed definitions and the relevant theoretical
expressions, In Section 3, an investigation of the possible
\underline{combined} effects on the two heavy vertex measurable
combinations of some models of New Physics will be performed, showing
that there  would be  distinct ``trajectories" in the ($\delta
R_b,\delta A_b $) plane in correspondence to different  models,
 and also a
brief discussion of some ``unnatural" possibility of light vertex-type
effects will be given, before drawing the final conclusions.
 A short Appendix
will be devoted to the derivation of some mass relationships in one of
the considered models, where one extra U(1) is involved.
\section{Definition of the second heavy quark vertex parameter}
 An immediate and natural way of defining a new heavy b vertex
parameter is to follow the philosophy that led to eq.(1) in the case
of    $\delta_{bV}$ and to introduce the quantity
$\eta_b$ as
\bq A_b = A_s(1+\eta_b)
\eq
i.e. as the ratio of the longitudinal polarization forward-backward
asymmetries for b and s-type quarks. The asymmetry $A_s$ (which
 corresponds
mathematically to that of practically massless b quarks) can be
written in a form similar to that of eq.(2) :\par
\bq A_s = 0.703(1 - 0.158( \Delta \kappa' + \delta'_s) - \Delta_{QCD}
{\alpha_s \over \pi} + ``negligible")  \eq
in which $\delta'_s$ is a vertex correction defined in [7]
 and $\Delta_{QCD}$ is a QCD factor of order one.
 With this
choice, one can easily see that the expression of $\eta_b$ becomes :
 \bq \eta_b = -\frac{2 (1- b^2)}{b( 1+b^2) }[
 \delta g^H_{Vb} -b \; \delta g^H_{Ab} ]
\eq
 The shifts $\delta g^H_{Vb,Ab}$ in eq.(11) take into account
 in the MSM the
effect of the would-be Goldstone exchange in the $Zb \bar b$ vertex
and also QCD effects due to the not negligible b-mass, whose complete
calculation has been given elsewhere [14] and that are, as such,
supposedly known. The important feature is that, \underline{in the
MSM} (but not a priori in the models of New Physics that we shall
consider) the effect on $\eta_b $
  of the charged would-be Goldstone boson (that is
proportional to $m^2_t$ in $ \delta_{bV}$ )
 is practically negligible, owing to the fact that
it gives the same contributions to $ \delta g_{Vb}$
 and to $\delta g_{Ab}$, that are nearly
cancelling in the combination of eq.(11). Thus, in the MSM prediction
for $A_b$, the ``heavy" b vertex component $\sim m^2_t$ can be
 ignored and the
relevant expression does only contain universal self-energies and
light vertices (and known QCD corrections). Obviously, this property is
a
priori  no longer verified as soon as one considers models of
 New Physics, for
which the relative role of $ \eta_b $ could be much more relevant or
fundamental.\par
To make the previous statement more illustrative, it is convenient to
reexpress the shifts of $ \delta_{bV}$ and $\eta_b$, rather than
in the ($g_V $, $g_A $) basis, in that provided by the (conventionally
defined ) ($ g_L $, $ g_R $) parameters. In that case, one can write:

 \bq \delta_{bV}
 = -\frac{4 (1+ b)}{( 1+b^2) } \left[ \delta g^H_{bL} -\frac{(1-b)}
{(1+b)}  \delta g^H_{bR} \right]
\eq

 \bq \eta_b = -\frac{2 (1- b)}{b } \left[ \delta g^H_{bR} +\frac
{(1-b) }{(1+    b)}  \delta g^H_{bL} \right]
\eq

As one sees, in the (L,R) basis the two shifts are orthogonal, which
means that effects that would not contribute one observable will be
revealed by the other one, and conversely.\par
To the previous remarks one can still add a property of $\eta_b $ that
is a direct consequence of our chosen definition eq.(9). In fact, if
one eliminates $ \delta g^H_{bL} $ in eq.(12), one obtains:

 \bq \eta_b = -\frac{2 (1- b)}{b } \left[ \frac{2(1+b^2)}{(1+b^2)}
 \delta g^H_{bR}-  \frac
{(1-b)(1+b^2) }{4(  1+    b)^2 }  \delta_{bV} \right]
\eq

and, to a very good approximation, this becomes:

 \bq \eta_b = -\left[ \delta g^H_{bR}- \frac
{1}{25}   \delta_{bV} \right]
\eq

showing that, once $ \delta_{bV} $ is experimentally known,
 the measurement
of $\eta_b $ fixes unambiguously the pure right-handed contributions
 from
various models to the ``heavy "  $ Zb\bar b $ vertex.\par

  After these  preliminary definitions, all the necessary
  ingredients to
formulate an unbiased search of New Physics effects in the ``heavy"
quark vertex sector are  at our disposal. One only has to take
eqs.(12), (13), insert a ``New Physics" apex to both the right and the
left-hand side,  and choose a  set of interesting
  models to be examined.  This will be done
in the forthcoming Section 3.\par
\section{Survey of models affecting the heavy b vertex}
The simplest known example of a model that contributes the heavy b
vertex is that with just   one extra Higgs doublet. In this case
 both the
charged and the neutral higgses will have to be considered. The charged
contribution  can be decomposed into
two terms. The first one essentially
reproduces that of the MSM (i.e.$ \sim \delta g_{bL} $ )  with the same
kind of $  m_t $ dependence (weighted by a factor $ \sim \cot ^2 \beta $
where $  \tan \beta $ is the ratio of the two VEV's ); the second one is
proportional to the product of $ m^2_b $ and $\tan^2 \beta $.
 As such, it
can only be relevant for very large values of $\tan \beta   \approx
m_t / m_b $.
Since it
   only modifies the right-handed $ Zb\bar b $ coupling,
it will generate a  suppressed effect   in  $\delta_{bV} $
(again,  of the same sign as that of the MSM ). More interestingly,
 it will
also  be able to affect
$ \eta_b $.
The neutral higgses sector is described by a larger set of parameters,
and
is therefore more model dependent than the charged one. In general,
 it will
affect both $\delta g_{bL} $ and $ \delta g_{bR} $ with terms
 proportional
to $ m ^2 _b $ and will consequently be only relevant if some
 enhancement
factor can be adjusted. In particular, this can be achieved when
 the value
of $ \tan \beta $ becomes very large. In this case, its contribution to
$ \delta_{bV} $ can be of opposite sign to that of the MSM [15].\par
These features of the simplest model with one extra Higgs doublet
 remain
essentially unchanged if one embeds it in a supersymmetric picture,
 with
the additional constraints between the various couplings and the
 existence
of other types of contributions to be taken into account.
This has been done
in great detail in a number of previous papers [16] for the
 specific case
of the so-called ``Minimal" Supersymmetric Standard Model
 (MSSM) [17] for
both small and large values of $ \tan \beta $. The results
 of all analyses
indicate that in some cases the effects of the Higgs sector and of the
genuine ``soft" supersymmetric sector can add up constructively,
 leading
to possible effects of a few percent that should be visible at future
measurements of $ \delta_{bV}, \eta_b $. \par
Among the configurations examined in ref.[16], that corresponding
 to large
$\tan \beta $ values was considered as a particularly interesting one.
The main motivation is that, while for small $\tan \beta $ values
the model  essentially
contributes $\delta g_{bL} $ but not $\delta g_{bR} $, in the  large
$\tan \beta $ case it can
affect both $\delta g_{bL} $ and $\delta g_{bR} $. As a consequence
of this,
 two independent experimental tests would become available
which would give rise to some implications. In particular, one
 would be able to
draw certain ``trajectories " in the $( \eta_b , \delta_{bV} ) $ plane
that would correspond to, or identify, a certain model and could be
experimentally ``seen ", at least in a certain part of the plane.\par
In the analyses of ref. [16], the contribution of the Higgs sector was
calculated using the SUSY  mass relationships valid at tree level
 in the
MSSM. Since it has become known [18] that these relationships are
appreciably
modified at one loop, one might be interested in evaluating
 the eventual
modification of the relevant trajectories (that are certain
 functions  of the
various higgses masses). Also, one might consider the effect of adding an
extra neutral Higgs to the model since this seems to be a reasonable
extension of the ``minimal" picture.\par
In this paper, we have examined the two possibilities and considered as
a tool model with one extra Higgs the so called $\eta $ model [19],
whose mass
relationships at tree level, that have been already examined in the
 literature
[20], show several interesting differences with those of the MSSM.
The results of our calculation will be only  shown for the Higgs sector
and for the related trajectories. The remaining contributions should be
identical with those computed in ref.[16] in the MSSM case.
 For the $\eta $
model, a separate calculation of non Higgs effects should be performed.
We believe, though, that the already existing  limits on the
mass of the
extra Z of this model, $ M_{Z'} > 500$ GeV [7], pushing
 the involved soft masses to large values,
  limit somehow in this model  their potential effect ( that should
not differ drastically, in any case, from the corresponding MSSM one).
\par
The relevant diagrams containing the various Higgses contributions are
shown in Fig.1; from these one derives compact   expressions that have
been already provided in the literature. Here we shall follow the
 notations
of Ref.[15] that, in the large $\tan \beta $ configuration chosen by us
 produce
the relatively simple formulae:

$$
 \delta g^H_{bR} =\frac{\alpha}{16 \pi s^2}\frac{ m_b^2 \tan^2 \beta}
{M_W^2} [ (1-\frac{4}{3}s^2) \; \rho_3[m_t,M_{H^+},m_t, M_Z]
$$

$$
- m_t^2 \; C_0[m_t,M_{H^+},m_t, M_Z]+
(s^2-c^2) \; \rho_4[M_{H^+},m_t,M_{H^+}, M_Z]
$$

$$
 + (-1/2+1/3 s^2) \; (\rho_3[m_b,M_A,m_b,M_Z]+\rho_3[m_b,M_h,m_b,M_Z] )
$$

\bq
 -\frac{1}{2}
\rho_4[M_h,m_b,M_A,M_Z]-\frac{1}{2} \rho_4[M_A,m_b,M_h,M_Z] ]
\eq

$$
 \delta g^H_{bL} = \frac{\alpha}{16 \pi s^2}\frac{ m_b^2 \tan^2 \beta}
{M_W^2} [+1/3 s^2 (\rho_3[m_b,M_A,m_b,M_Z]+\rho_3[m_b,M_h,m_b,M_Z])
$$

\bq
-\frac{1}{2} \rho_4[M_h,m_b,M_A,M_Z]- \frac{1}{2} \rho_4[M_A,m_b,M_h,M_Z] ]
\eq

Here $\rho_{3,4}[m_1,m_2,m_3,M_Z]$  and $C_0[m_1,m_2,m_3,M_Z]$
are the functions introduced  in the appendix of
ref.[15 ]. The masses that appear in
the previous expressions are those of the charged Higgs
$ ( M_{H^+} ) $ , of the CP-odd neutral Higgs $ ( M_A ) $ and of that
CP-even neutral Higgs $ ( M_h )$
    whose mass is nearly degenerate with  $ M_A $ in
the MSSM and in the $ \eta $ model.Starting from the given expressions,
one only has to insert, at a certain level of accuracy,
 the mass relationships
of the various  models that are, in general, not the same.
In particular,
the famous tree-level
formulae of the MSSM
and the corresponding ones  of the $ \eta $ model [20] can
 be substantially
different. For example, one finds in the first case the equality :
\bq M^2_{H^+}= M^2_A+ M^2_W \eq
whilst in the second model one has:
\bq M^2_{H^+}= M^2_A+ M^2_W \left[1- \frac{2 \lambda^2}{g^2}
 \right] \eq
where $ \lambda $ is a free  parameter. Also, one finds a bound
for the lightest neutral in the MSSM, that becomes sensibly larger
in the other case [20]. At one loop,extra  not negligible differences
  can arise in
both models, which could in principle give rise to observable
effects.\par
Motivated by the previous argument, we have calculated eqs.(16),
(17) inserting
the one-loop mass relationships of the two models.For the MSSM,
 these are
known and can be found in the literature [18]. For the
 $ \eta $ model,in
the chosen configuration, they are given in the short Appendix. The
numerical values of $ \delta_{bV} $ and $ \eta_b $ are shown in the
following Figures. They will depend on $ m_t $ (from the charged
sector),
on $ m_b \tan \beta $ (from both sectors ) and on \underline{one}
 residual
neutral mass chosen to be $ M_A  \approx  M_h $ .The value
 of $ M_{H^+} $
remains fixed by the choice of the configuration, as shown in
 Appendix ,
for the MSSM.In the case of the $\eta $ model,for which extra
 parameters
exist, we  have chosen the situation that optimizes the effect
 and thus
the related figures are actally showing the maximal deviations
 that the
model can produce.
All the numerical results are given for $ m_b $ = 5 GeV,
 $ \tan \beta $ =
70, following the approach of Ref.[16].

To get a qualitative
feeling   of the differences obtained
by using the modified mass relationships, we show in Figs.2, 3 the
trajectories corresponding to the MSSM with mass constraints
 at tree level,
eq.(18), and at one loop. One sees that one  effect is that
 of ``smoothening"
the $m_t$ dependence, particularly in the heavy
mass region,say, between 150 GeV and 200
GeV (intermediate and  upper lines)(this is a
consequence of the fact that in the charged Higgs
contribution this dependence is now  weakened in the relevant
 ratio between
the top and the Higgs masses ). Also,  one notices a systematic
 (small) decrease in $\eta_b$, compensated by a corresponding
(small) increase in $\delta_{bV}$.\par
In fact, the compensation between $\eta_b$ and $\delta_{bV}$ is
 quite general,
in the sense that for small $M_A$ values the full (positive )
 effect is on
the second parameter, while for large $M_A$ only the first one
 is modified.
This is related to the fact that $\eta_b$ is dominated by right-handed
effects, that are peculiar of the charged Higgs contribution whose
 decoupling
is slower than that of the neutral ones (that give the important effect
on $\delta_{bV}$ ).\par
If we accept the experimental available indications [8] that seem
 to prefer
\underline{positive} (or, at least, not too negative ) $\delta_{bV}$
shifts,  we
conclude that the most relevant  part of the Higgs sector
 trajectory of this model lies in the positive $\eta_b$ region of the
 plane
( with the exception of the fraction that would correspond to
 substantial
$\delta_{bV}$ effects (larger than,say, two percent) i.e. to very
 small $M_A$
values , where the shift on $\eta_b$ could be negative).
Since the same feature seems to be valid for the remaining
genuinely supersymmetric contributions of the model [16],
we conclude that the simultaneous observation of (small) positive
 deviations
in either $\delta_{bV}$ or $\eta_b$, or possibly in both ,could be
interpreted
as the experimental evidence for this model in the
considered region of its parameter space. This would require a
precision of the two measurements of the order of a relative one
 percent,
although in certain favourable cases the shifts could be
larger than that, particularly if the effects from the Higgs and the
genuine SUSY sector
added in a substantial way as they seem to be willing to do.\par
The case of the $\eta$ model is illustrated in Fig.4, only showing
 the
situation where the mass constraints are used at one loop.
As one sees, the results for the Higgs sector are very similar
to those of the previous example,with a small general increase
of $\eta_b$ and practically no change in $\delta_{bV}$. Since
we expect that other contributions are somehow depressed in this
case , we would conclude that the trajectories of this model
are qualitatively similar to those of the MSSM (with possibly
smaller overall effects); in other words, the presence
of one more neutral scalar does not affect the
trajectory in this case.  Whether this is a general feature
of SUSY models with one extra (singlet) scalar remains to be
 investigated; we
postpone the discussion of this point to a next forthcoming paper.\par
It can be interesting to remark that in the ``orthogonal" case
of Technicolor-type modifications of the MSM, the associated
trajectories would be completely different for a wide class of
models.This can be deduced from the analysis presented in
reference [21] where the contributions to
$\delta_{bV}$ were computed. In fact, for a class of ``walking
technicolor" cases the effect on $\delta_{bV}$ was negative and
of purely left-handed type, leading in any case to negative corrections
to $\eta_b$ as one can easily verify from the defining eqs.(12), (13).
The exception to this statement would be represented by a class of
special models where fermion masses are due to the presence and mixing
of technibaryons [22],that produce positive shifts in $\delta_{bV}$.
But
for these models, the shift in $\eta_b$ can be written  to a good
approximation, using again eqs.(12), (13) as follows:
\bq \eta_b \simeq \delta_{bV} \frac{1-5 c^2}{5(5+c^2)}
\eq
where $c^2=  \sin^2 \alpha / \sin^2 \beta'$ and $\alpha$, $\beta'$
 are the
two mixing angles of the model.Varying this ratio from zero to
 infinity
fixes $\eta_b$ in a region between ,practically, zero
 and -$\delta_{bV}$
as shown in the next Fig.5. Thus, the observation of two small
effects of opposite sign with a negative $\eta_b$ would provide
 a rather
peculiar evidence for this special model.\par
To conclude our investigation, we have considered the
 (less attractive, in
our opinion) possibility that the origin of small discrepancies
 in ${\delta R_b \over R_b}$
and ${\delta A_b \over A_b}$ is due to effects of light-fermion type. Firstly,
we
have considered the class of models with one extra Z' of $ E_6$
 origin that
has been often considered in the literature [19]. For these models,
 strong
experimental constraints on the mixing angle exist [7] that limit its
modulus to be less than, say, one percent. Using this extreme value
 as the
tolerated limit for every single model (which is somehow optimistic)
we obtain the effects shown in Fig.6. As expected,
the possible effects of this kind are always below the one
percent level and are spread in the  (${\delta R_b \over R_b}$
, ${\delta A_b \over A_b}$)  plane.
In other words, the existing limits on the mixing angle seem to
prevent interesting effects from these models. Note, accidentally,
that the contribution coming from the $\eta$ model
(that would belong in the chosen configuration of large $\tan\beta$
values to positive mixing angles) goes in the opposite direction
to that of the Higgs sector,which represents a negative feature of
the model. We have repeated our analysis for an extra $Z'$
predicted by Left-Right symmetry models and for higher vector
bosons predicted by various types of different models,in particular
compositeness inspired models ($Y$, $Y_L$, $Z^*$) [24] and
alternative symmetry breaking models ($Z_V$) [25]
.  As in the first case, the
limits imposed by precision tests in the light fermion sector
prevent from getting large effect on $R_b$ and $A_b$ as one can see
in Fig.6.\par
We can summarize the results of this preliminary investigation
as follows.
Assuming as a realistic goal a final experimental accuracy on the
measurements of both $A_b$ and $R_b$ of a relative
 one percent,
the best chances of providing visible signals seem to belong to
 models of
New Physics that can affect the ``heavy" b-vertex
component. Among these, we have seen that those of SUSY
 type are associated
to trajectories in the plane of the variations of $\delta_{bV}$ and
$\eta_b$ that differ substantially from those of technicolor type. We
stress the fact that this differentiation is made possible by the
\underline{combined} measurements of the two observables; for instance,
the discovery of a positive effect in $\delta_{bV}$ could not
discriminate the models of Figs.2, 3, 4 from that of Fig.5.
Should this effect (that is apparently not disallowed by the existing
 data)
survive in the future, the role  of a high precision measurement
of $\eta_b$ would become, least to say, fundamental.\par
Before concluding this paper we would like to make a rather speculative
remark concerning the possibility that a positive shift
 of $R_b$
is observed with no effect on $A_b$. From a purely technical
 point of view,
it might be possible to explain this effect in a picture where the MSM
calculation is still valid, but where the \underline{effective} axial coupling
of
Z to the \underline{top} is slightly decreased. In fact, in the large
$m_t$ limit, the dominant contribution to $\delta g_{bL}$ can be
 expressed
in the form:
 \bq    \delta g_{bL} \simeq \frac{\alpha}{8 \pi s^2}
 \frac{m^2_t}{M^2_W}
g_{t,A}  \eq
and values of $g_{t,A}$ slighty smaller than one-half (with no effect
on the corresponding b-vertex ) could provide this possible
 deviation,
thus motivating searches of reasonable  models where the axial
 ``form factors" of heavy
quarks can be possibly modified [23].

{\Large \bf Acknowlegements }\par
One of us (C.V.) acknowledges some useful discussions with B.W.Lynn and
H.Schwarz during a stay at UCLA in summer 1993.

\newpage

{\Large \bf Appendix}\par
In this Appendix  we give the expressions of
the relevant radiative corrections (R.C.) to   eq.(18)
in the MSSM  and to eq.(19)  in the $\eta$ model.
The Higgs sector of the MSSM at tree level is described by two
 parameters,
$\tan \beta$ and $M_A$; when we include the radiative corrections
  all the
parameters which describe the spectrum of the theory enter in the mass
 formulae.
 The most important contributions to the  R.C.   come from the
stop-sbottom sector, so we must fix:
the soft squark masses ($m_{\tilde{t}_L,R}=m_{\tilde{b}_L,R}=m_{\tilde{q}}
\simeq 1     $  TeV);
the trilinear SUSY  breaking parameters
($A_t=A_b= 100$ GeV) ; the SUSY $H_1 H_2$ coupling  $\mu$ and  of course
the top mass.
In the large $\tan \beta$ limit the one loop mass relationships  read [18]:

\bq M^2_{H^+}= M^2_A+M^2_W+ \Delta M^2_{H^+}
\eq
  where
$$
 \Delta M^2_{H^+}=\frac{3 g^2}{32 \pi^2 M^2_W} [ 2 m_t^2 m_b^2
 \tan^2 \beta
- M^2_W ( m_t^2+ m_b^2 \tan^2 \beta )
$$

$$
+\frac{2}{3} M_W^4 ] \log{ \frac{m_{\tilde{q}}^2}
{m_t^2}}+ \frac{3 g^2}{96 \pi^2}  [ m_t^2 \left( \frac{\mu^2-2 A_t^2)}
{m^2_{\tilde{q}}} \right)
$$

$$
 + m_b^2 \tan^2 \beta  \left( \frac{\mu^2-2 A_b^2)}
{m^2_{\tilde{q}}} \right)  ] +
\frac{3 g^2}{64 \pi^2} M^2_W [ \frac{m_t^2 m_b^2 \tan^2 \beta}{M^4_W}
\left( \frac{A_t+ A_b}{m_{\tilde{q}}^2} \right)^2
$$

\bq
- \frac{\mu^2}
{m_{\tilde{q}}^2} \left( \frac{m_t^2+ m_b^2 \tan^2 \beta}{M^2_W}
   \right)^2 ]
- \frac{ 3 g^2 m_t^2 m_b^2 \tan^2 \beta}{192 \pi^2 M_W^2}
\left( \frac{ A_t A_b-  \mu^2}{m_{\tilde{q}}^2}   \right)^2
\eq

The radiatively corrected  mass $M_h$
  of the  CP-even neutral Higgs which runs into the loop of Fig.1
is always nearly  equal to $M_A$ .

In the $\eta $ model the tree level Higgs sector is defined by
 4 parameters:
$\tan \beta, M_A, x, \lambda$.
The new parameter $x$ is the VEV\def\bq{\begin{equation}}
 of the extra complex Higgs field N  and  fixes
the scale of the breaking of the extra U(1) gauge group, so  naturally
$x \gg  v_1, v_2$.
In this large $x$ limit the Higgs sector, that is described by 3 CP-even,1
CP-odd and 1 charged-state, effectively  reduces, at the $M_Z$  scale, to that
of the MSSM with the following identifications:
$\mu=\lambda x, m_3^2=\lambda A_{\lambda} x$ ($m_3^2$ is the soft SUSY
breaking term of the operators
$H_1H_2$ in the MSSM, $A_{\lambda}$ is the trilinear soft term
 which multiplies
the product $N H_1 H_2$ in the potential).
When  R.C. are evaluated, besides the parameters of the MSSM
there is another Yukawa coupling $h_E$ of the exotic
quark sector ($m_{\tilde{E}}=m_{\tilde{q}}, \;  A_E=A_t$)
So finally  the extra new parameters are $\lambda,\;x $ and $h_E$.
We fix  $x$  via the mass of the extra $Z'$ boson:
$M_{Z'}= 25/18 g_1^2 x^2= $0(1 TeV).
The exotic Yukawa coupling gives very little contributions (some GeV)
 to the
``standard" Higgs  sector
and  can be safely  fixed to 1.
The Higgs spectrum is at the contrary very sensitive to the $\lambda$
parameter: this strong dependence is
exibited by  the charged Higgs sector (see eq.(19))
and by the ligthest CP-even mass.
As shown in ref.[20] (for values of $M_A< M_{Z'}$ ) the ligthest
 Higgs mass
($M_l$) is a
convex parabola in the $\lambda^2$- $M_l$ plain.
The imposition of the experimental bound   $M_l \geq 60$
GeV gives a very strong upper limit on $\lambda$ (tipically
 $\lambda<0.4$);
 therefore the difference between the charged Higgs mass (for
fixed $M_A$) in the
two models cannot be  arbitrarily large.
The mass $M_h$ is again  nearly equal to $M_A$.
So, the only effective difference between the $\eta$
 model and the MSSM, in this region of the
space parameters, is contained in the relation $M_{H^+}\!-M_A$:
\bq M^2_{H^+}= M^2_A+M^2_W (1- \frac{2 \lambda^2}{g^2})+
\Delta M^2_{H^+}+
\Delta' M^2_{H^+}
\eq
where $\Delta M^2_{H^+}$  is the same as in  eq.(20)
 with the suitable identifications and
 $\Delta' M^2_{H^+}$  is the small contribution of the exotic sector:
\bq\Delta' M^2_{H^+}=-\frac{3}{8 \pi^2} M^2_W
\frac{\lambda^2}{g^2} h_E^2 \left[
\log{\frac{m_{\tilde{q}}^2+ m_E^2}{M_Z^2}}
-\frac{1}{6}\frac{ A_E^2 m_E^2}{m^2_{\tilde{q}}+ m_E^2} \right]
\eq
In general when $\lambda \rightarrow 0$  we have  the
  same relationships
$M_{H^+}\!-M_A$ as in the MSSM and the trajectories in the plane
($\delta_{bV}, \eta_b$)  are the same.
What we have shown in Fig.4
are the trajectories with   the maximum  value  of $\lambda$   such
that  the neutral Higgs sector is beyond  the present experimental
 bound.\\

\newpage

\par
\centerline { {\bf Figure Captions }}\par
\begin{description}
\item[Fig.1 :]  Self energy and vertex corrections to
 the $Z b \bar b$  vertex

\item[Fig. 2 :]  Plot in the ($\delta_{bV}, \eta_b)$ plane of
 the corrections
 (in percent)
in the MSSM case  with the relationships $M_{H^+}\!-M_A$ at tree
level (see eq.
(18)). There  are 16 point for  each  ``curve" ,each one  corresponding
to a given value  of $M_A$,in particular ( starting from the right  to
the left):   $M_A$= 40, 45, 50, 55, 60, 65, 70, 75, 80, 90, 100,
 120, 140, 160, 180, 200 GeV.
The upper line corresponds to  $m_t$=200 GeV, the intermediate
 one to  $ m_t$=
150 GeV  and  the lowest one to  $m_t$= 110 GeV.

\item[Fig. 3 :] The same  as before  for the MSSM but with the mass
  relationships
at one loop (see eq.(22)).

\item[Fig. 4 :] The same as before   but for the $\eta$  model
   and with  the
mass relationships at  one loop (see eq.(24)).

\item[Fig. 5 :] The set of allowed trajectories for the Kaplan model
discussed in ref.[21,22]
at variable ratio $c^2$ of the two mixing angles.

\item[Fig. 6 :]Maximal allowed $Z-Z'$ mixing
  effects in the (${\delta R_b \over R_b}$, ${\delta A_b \over A_b}$)
 plane, from $E_6$ based models with $-1 \leq \cos(\beta) \leq
+1 $ (dashed), from L-R symmetry based
models with $\sqrt{{2 \over 3}} \leq \alpha_{LR} \leq \sqrt{2}$ (full), in
both cases with $|\theta_M| = 0.01$. We have also indicated the
trajectories or small domains allowed for various alternative models
of higher vector bosons ($Y$, $Y_L$, $Z^*$, $Z_V$)
taking into account the constraints established in ref.[7].

\end{description}

\end{document}